\documentclass[11pt,a4paper]{article}
\usepackage{jinstpub}

\def\Momatom{{\overline{\Gamma}_a}}

\def\kv{{\bf k}}

\def\pv{{\bf p}}

\def\rv{{\bf r}}

\def\vv{{\bf v}}

\def\Ev{{\bf E}}
\def\Fv{{\bf F}}

\def\nv{{\bf n}}

\def\be{\begin{equation}}
\def\ee{\end{equation}}
\def\ni{\noindent}

\def\lambdabar{\lambda\raise0.4ex\hbox{\kern-0.5em\hbox{--}}\ }
\def\lambdaC{\lambda\raise0.5ex\hbox{\kern-0.5em\hbox{--}}_{\rm C}}
\def\lambdabarc{\lambda\raise0.5ex\hbox{\kern-0.4em\hbox{--}}_{\rm c}}
\def\lambdabarE{\lambda\raise0.5ex\hbox{\kern-0.4em\hbox{--}}_{\rm e}}
\def\lesssim{\,{\lower0.5ex\hbox{$\stackrel{<}{\sim}$}}\,}
\def\gtrsim{\,{\lower0.5ex\hbox{$\stackrel{>}{\sim}$}}\,}

\def\appvt{{\bf u}}
\def\lc{L_{\rm coh}}
\def\td{\tau} 

\def\t{_{\perp}}

\def\T{_{\rm T}}
\def\L{_{\rm L}}

\def\epol{{\bf{\hat e}}}
\def\ccdot{\!\cdot\!}

\def\eg{{\it e.g.}}
\def\ie{{\it i.e.}}
\def\arad{{\bf a}_{\rm rad}}

\title{\boldmath 
Classical spectral sum rules and ``half-naked'' electron effects in radiation from relativistic electrons in external field}
 \author{X. Artru}
\affiliation{Universit\'e de Lyon, Institut de Physique des deux Infinis (IP2I Lyon),  \\
Universit\'e Lyon~1 and CNRS, France}
\emailAdd{x.artru@ipnl.in2p3.fr}

\abstract{ 
Two properties of the radiation emitted by a relativistic electron in an external field, in the classical approximation, are presented in details:
1) spectral sum rules and their relationship with the sum rules for oscillator strength in atomic physics. \textit{Filtered} sum rules can be used to remove infrared or ultraviolet divergences.  2) ``half-naked electron'' effects like the Landau-Pomeranchuk-Migdal and Ternovskij-Shul'ga-Fomin effects. These properties are best understood when introducing the {\it apparent transverse velocity}.}  

\keywords{Only keywords from JINST's keywords list please}

\begin{document}
\maketitle
\flushbottom

\section{Introduction}

Radiation by relativistic electrons in an external field should, in principle, be treated with Quantum Electrodynamics (QED). If the external field is enough weak and smooth, 
the electron motion is almost classical and the photon momenta are negligible compared to the electron one, so that the trajectory is practically unchanged at each photon emission. 
The electron can therefore be considered as a classical current and one can use the 
classical theory of radiation. This is the case of synchrotron and undulator radiation. 
This classical approach fails when the external field {\it in the instantaneous electron frame} reaches the order of the critical QED field $E_{\rm c} =  m^2/e \simeq 1.32~10^{18}$ V/m or varies significantly within a time $\lambdabarE = 1/m \simeq 386$ fm. %
\footnote{
We work with the natural unit system where $c=\hbar=1$ $\,\alpha=e^2/(4\pi)\simeq1/137$. The electron energy is $\epsilon= \gamma m = m \, (1-\vv^2)^{-1/2}\gg m$. The letter "$\gamma$" can also designate a photon.
The photon momentum is $\kv=\omega\,\nv$. }
In that case, the photon can take a non-negligible fraction of the electron energy. However a semiclassical formula which takes into account the \textit{recoil effect}, together with a dependence on the electron spin, has been proposed by  Ba\"ier and Katkov \cite{B-K}. 
Its domain of validity is discussed in Ref. \cite{XA-Ischia}, where a new derivation is presented for the case of a plane wave external field.

In this paper, we consider the case of weak and smooth field and derive some classical properties of the radiation emitted by relativistic electrons in external field: coherence length, ultraviolet cutoff and sum rules. They apply in particular to synchrotron radiation, undulator radiation and, for electrons in the energy range $\sim[10^2-10^3]$ MeV, to channeling radiation and coherent bremsstrahlung (CBR). We also consider the coherence length effects in the low-frequency part of ordinary bremsstrahlung, where the classical theory applies, and revisit the Landau-Pomeranchuk-Migdal and Ternovskij-Shulg'a-Fomin effects. 


\section{Classical radiation formula}

For a selected photon polarization $\epol$, with $\epol \!\cdot\! \epol^*=1$, the classical energy-angular spectrum of the radiation writes
\be\label{cov1} 
dW/d^3\kv = \omega dN(\epol)/d^3\kv =(16\pi^3)^{-1} \,  \left| \epol^*\ccdot\arad(\kv)\right|^2 \,,
\ee
where $\arad(\kv)$ is the 3-dimensional Fourier transform of the {\it radiated} electric field, $\Ev_{\rm retarded} - \Ev_{\rm advanced}$, and is classically given by (see, \eg, Refs. \cite{JACKSON,RAD})%
\footnote{An alternative derivation of  $\arad(\kv)$ using only energy conservation can be found in Ref. \cite{XA-spont-stim}.}
\be\label{noncov} 
 \arad(\kv)  = e \int_{-\infty}^\infty d\td \, \exp(i\omega\td) 
 \, \appvt(\td) \,. 
\ee
$\td=t-\nv\ccdot\rv$ is the \emph{detection time} up to a constant; $\nv=\kv/|\kv|$; 
$\omega\td=k_\mu X^\mu\equiv\phi$ is the {\it local emission phase}; $\appvt(\td) \equiv {d\rv_\perp}/{d\td}$ is the {\it apparent transverse velocity}, related to the real one by
\be
 \appvt(\td)= (dt/d\td ) \, \vv_\perp(t) = \vv_\perp(t) / [1-\nv\cdot\vv(t)] \,.
\ee
The subscript $\perp$ means perpendicular to $\nv$.
An alternative formula,
\be\label{noncov'} 
 \arad(\kv)  
=   \frac{ie}{\omega} \int_{-\infty}^\infty d\td \, \exp(i\omega\td) \,  \frac{d^2\rv_\perp}{d\td^2}  \,,
\ee
involves the {\it apparent transverse acceleration}. 

\paragraph{Ultrarelativistic approximation and coherence length.}
In the following we assume $\gamma\gg1$, \ie $|\vv|\simeq1$, and small deflection angles: $|\vv(t)-\vv(t')| \ll1$. Most photons are emitted at small angle from the trajectory. The other ones are much softer and will be neglected. Then, $|\vv\t(t)|\ll1$ and
\be \label{dtau}
d\td/dt 
\simeq [\gamma^{-2}+\vv\t^2(t)]/2  \ll 1\,,
\ee
\be \label{appvt}
\appvt(\td) 
\simeq 2 \vv\t / [\gamma^{-2}+\vv\t^2(t)] \,.
\ee
 $|\appvt|$ is plotted as function of $|\vv\t|$ in Fig. 2. It reaches a maximum, equal to $\gamma$, at $|\vv\t|=\gamma^{-1}$. 
The {\it coherence length}, defined as the distance over which $\phi=\omega\td$ changes by 1 radian, is given by  
\be \label{Lcoh}  
\lc(\omega,\gamma,v_\perp)  = dt/d\phi 
\simeq (\lambda/\pi) \, (\gamma^{-2}+\vv\t^2)^{-1} 
\ee
for a straight piece of trajectory. It can reach macroscopic values.

 \paragraph{Ultraviolet cutoff.}
$\arad(\kv)$ becomes negligible when the apparent motion is nearly uniform within a coherence length. 
This leads to a classical cutoff $\omega_{\rm c,class} \sim \gamma^3/R$ for a circular motion or
$2\pi\gamma^2/\lambda$ for the ``inverse Compton effect''.
If $\omega_{\rm c,class} / \epsilon$ is not small, the classical theory fails and, taking into account the recoil effect, $ \omega_{\rm c,class}$ is replaced by
\be
\omega_{\rm c, quant} =  \omega_{\rm c,class} \,\epsilon / (\epsilon + \omega_{\rm c,class}) \,.
\ee

 \section{Spectral sum rules \cite{RAD}}  
At fixed $\nv$, we define the $p^{\rm th}$ momentum $M^{(p)}$ by
\be \label{moment-class}
\int_0^\infty d\omega \, \omega^{p} \, \frac{dN(\epol)}{d\omega d^2\nv} = M^{(p)}(\epol,\nv) \,.
\ee
For {\it linear} polarization (real $\epol$), we have
\begin{subequations} \label{sumrule} 
\begin{align}
M^{(-3)}(\epol,\nv) &= \frac{\alpha}{4\pi} \int d\td \, (\epol\ccdot\rv\t)^2 \,, 
\\
M^{(-1)}(\epol,\nv) &= \frac{\alpha}{4\pi} \int d\td \, (\epol\ccdot d\rv\t/d\td)^2 \,,
\\
M^{(1)}(\epol,\nv) &= \frac{\alpha}{4\pi}  \int d\td \, (\epol\ccdot d^2\rv\t/d\td^2)^2 \,,
\end{align}
\end{subequations}
obtained from Eq. (\ref{noncov'}) using the Parseval-Plancherel formula.
Integrating $M^{(1)}$ over $\nv$ and summing over $\epol$ gives the \emph{Li\'enard formula} for the radiated energy
\be \label{Larmor}
W=\frac{2\alpha}{3m^2} \int dt (F\L^2 + \gamma^2 \Fv\T^2) \,,
\ee
$F\L$ and $\Fv\T$ being the components parallel and perpendicular to $\vv(t)$ of the Lorentz force $\Fv[\rv(t)]$ acting on the electron. 
 Eq. (\ref{Larmor}) has something to tell about coherent bremsstrahlung~: 
 the total CBR energy on a family of atomic planes does not depend on the angle $\psi$ between the incident particle momentum and the planes, as observed in Ref. \cite{GOUANERE}. 
%

The right-hand sides of Eq. (\ref{sumrule}a-c) are infinite if the integrands do not vanish at $t\to\pm\infty$.
Let us consider, however, a motion of the form 
\be \label{quasi-uniform}  
\rv(t)=\rv_0+ \bar\vv t +\delta\rv(t) \,,
\ee
$\delta\rv(t)$ being a periodic or pseudo-periodic%
\footnote{an example of pseudo-periodic motion is the \textit{rosette} motion in axial channeling.}
fluctuation about the uniform motion.  
For a given $\nv$,  $\rv\t(t)$ can be re-parametrized in terms of $\tau$ as
\be \label{quasi-uniform-apparent}
\rv\t(\td)=\bar\rv(\nv)+ {\bf \bar u}\t  \td +\delta\rv\t(\nv,\td)\,,
\ee
where $\bar\rv(\nv)$ is chosen such that $\langle \delta\rv\t(\nv,\td) \rangle = {\bf0}$. Then  
we can replace $\rv\t(\td)$ by $\delta\rv\t(\nv,\td)$ 
and $d\rv\t/d\td$ by $ \delta{\bf u}\t = d\delta\rv\t/d\td $ 
in Eqs. (\ref{sumrule}). That eliminates spurious zero-frequency contributions to the moments $M^{(p)}$. 
Then the right-hand sides of Eqs. (\ref{sumrule}a-c) diverge only linearly and one can define the {\it average derivative in the detection time}, 
\begin{subequations} \label{dsumrule}
\begin{align}
\langle dM^{(-3)}(\epol,\nv)/d\td \rangle &= \frac{\alpha}{4\pi}\,
\langle [\epol\ccdot\delta\rv\t(\nv,\td)]^2  \rangle \,, 
\\
\langle dM^{(-1)}(\epol,\nv)/d\td \rangle &= \frac{\alpha}{4\pi}\,
\langle (\epol\ccdot \delta{\bf u}\t(\nv,\td))^2 \rangle \,,
\\
\langle dM^{(1)}(\epol,\nv)/d\td \rangle &= \frac{\alpha}{4\pi} \, 
\langle [\epol\ccdot d\delta{\bf u}\t(\nv,\td)/d\td]^2 \rangle\,.
\end{align}
\end{subequations}
Eqs. (\ref{sumrule}), (\ref{Larmor}) or (\ref{dsumrule}) can be applied to undulator and channeling radiations. 

A familiar case where Eq. (\ref{quasi-uniform}) does not hold is single- or multiple scattering with $\vv(+\infty)\ne\vv(-\infty)$. Then $M^{(-1)}$ and $M^{(-3)}$ are infrared divergent. Besides, $M^{(1)}$ is ultraviolet divergent if scatterings are considered as breaks of the trajectory. In fact, $\omega_{\rm c,class} = \infty$
in this case. 
If we replace the breaks by continuous bends at atomic scale $a\sim0.5$ \AA, we have  $\omega_{\rm c,class} 
\sim \gamma^2/a\gtrsim \epsilon$ and Eq. (\ref{Larmor}) would overestimate the bremsstrahlung power, the right formula being $dW/dt=\epsilon/X_0$. However the classical formula, Eq. (\ref{noncov}) is still valid for the low frequency part of bremsstrahlung.

Ultraviolet or infrared divergences can be removed in \textit{filtered sum rules} 
\cite{XA-Elbruz}. These involve the convolution of the trajectory by a function $f(\td)$. For example, with $p=-1$, 
\be  \label{filter}
\int_0^\infty d\omega \, \omega^{-1} \, |\tilde f(\omega)|^2 \, \frac{dN(\epol)}{d\omega d^2\nv} =
\frac{\alpha}{4\pi} \int d\td \,  ({\bf u} \otimes f)^2 \,,
\ee
where the {\it filter} $\tilde f(\omega)$ is the Fourier transform  of $f(\td)$. If $|\tilde f(\omega)|^2$ is a high-pass filter, the integrals of Eq. (\ref{filter}) are infrared convergent.
Eq. (\ref{filter}) has been used to eliminate spurious infrared divergences in the simulation code of channeling radiation of Ref.  \cite{XA-code}. 

For {\it circular} polarizations $\epol^\pm$ corresponding to helicity $\pm1$, we have 
%
\begin{subequations} \label{sumrule-circ}
\begin{align}
M^{(-2)}(\epol^+,\nv)  - M^{(-2)}(\epol^-,\nv) 
&= \frac{\alpha}{4\pi} \int d\td \, [\rv\t \times \appvt] \,, 
\\
M^{(0)}(\epol^+,\nv)  - M^{(0)}(\epol^-,\nv) 
&= \frac{\alpha}{4\pi} \int d\td \, [\appvt \times d\appvt/d\td] \,.
\end{align}
\end{subequations}
In the condition of Eq. (\ref{quasi-uniform}), one can replace $\rv_\perp$ and ${\bf u}\t$ by $\delta\rv_\perp$ and $\delta{\bf u}\t$.

\paragraph{Analogue in atomic spectroscopy}
(this analogy has been studied in collaboration with Patrice Garnier  \cite{Garnier}).
The radiative width of the state $|a \rangle$ of a one-electron atom can be written as
%
\be
\Gamma_a=  \sum_\epol \int d^2\nv \, \Gamma_a(\nv,\epol) \,,
\ee
$\nv$ and $\epol$ being the photon direction and polarization. In the dipole approximation, 
%
\be
\Gamma_a(\nv,\epol) = \frac{\alpha}{2\pi}  \sum_b \Theta(E_a-E_b) \, (E_a-E_b)^{3} \, | \langle b| \epol^*\ccdot\rv |a \rangle |^2 \,,%
\ee
$\Theta(x)$ being the Heaviside function. 
Let us define  the ``$\Gamma_a$'' and ``$\Momatom$'' moments: 
\begin{subequations} \label{moment-quant}
\begin{align}
\Gamma_a^{(p)}(\nv,\epol)   &= \frac{\alpha}{2\pi} \sum_{b}  (E_a-E_b)^{p+3} \, 
| \langle b| \epol^*\ccdot\rv |a \rangle |^2 \, \Theta(E_a-E_b) \,,
\\
\Momatom^{(p)}(\nv,\epol)   &= \frac{\alpha}{4\pi} \sum_{b}  (E_a-E_b)^{p+3} \, | \langle b| \epol^*\ccdot\rv |a \rangle |^2 \,.
\end{align}
\end{subequations}
$\Momatom^{(p)}$, which has no $\Theta$ function but an extra factor 1/2, obeys the linearly polarized sum rules for oscillator strength \cite{Bethe}: 
\begin{subequations} \label{sumrule-atom} 
\begin{align}
\Momatom^{(-3)}(\nv,\epol) &=   \frac{\alpha}{4\pi} \langle a| (\epol\ccdot\rv)^2 |a \rangle \,, \\
\Momatom^{(-1)}(\nv,\epol) &= \frac{\alpha}{4\pi} \langle a|(\epol\ccdot\pv)^2 |a \rangle /m^2 \,,\\
\Momatom^{(1)}(\nv,\epol) &= \frac{\alpha}{4\pi} \langle a| (\epol\ccdot\nabla V)^2 |a \rangle /m^2 \,.
\end{align}
\end{subequations}
If $|a \rangle$ is a Rydberg state  (large $n$ and $l$), we have $\Momatom^{(p)} \simeq \Gamma_a^{(p)}$ 
for odd $p$, up to corrections of order $\hbar$. Indeed, 
 $\langle b| \epol^*\ccdot\rv |a \rangle$ is important only for small $|E_a-E_b|$ and, for odd $p$, 
states of positive and negative $E_a-E_b$ contribute almost symmetrically to Eq. (\ref{moment-quant}b). 
Thus, with the substitutions $\Momatom \Rightarrow \Gamma_a$,~ $\pv/m=d\rv/dt$,  $\nabla V/m = -d^2\rv/dt^2$ and $t\Rightarrow\td$, Eqs. (\ref{sumrule-atom}) appear as a particular case of Eqs. (\ref{dsumrule}).
Instead, for even $p$, we have $\Gamma_a^{(p)} \gg \Momatom^{(p)}$, the latter being $\sim {\cal O}(\hbar)$, as in the Thomas-Reiche-Kuhn sum rule.   

For circular polarizations we have \cite{Klar} 
\begin{subequations} \label{sumrule-atom-circ} 
\begin{align}
\Momatom^{(-2)}(\epol^+,\nv)  - \Momatom^{(-2)}(\epol^-,\nv) 
&= \frac{\alpha}{4\pi}\langle a| {\bf L}\ccdot\nv |a \rangle /m \,,
\\
\Momatom^{(0)}(\epol^+,\nv)  - \Momatom^{(0)}(\epol^-,\nv)  
&= \frac{\alpha}{4\pi} \langle a| [\pv\times\nabla V]\ccdot\nv |a \rangle /m^2 \,,
\end{align}
\end{subequations}
which, due to $\Delta\Gamma^{(p)}\simeq\Delta\Momatom^{(p)}$ for an even $p$ for Rydberg state, is a particular case of Eqs. (\ref{sumrule-circ}).

 \section{``Half-naked'' electron effects.} 
 
 An ultrarelativistic electron in uniform motion is accompanied by its Lorentz-contracted Coulomb field, which ressembles a radiation field and whose spatial Fourier transform is   
%
\be\label{WW} 
\tilde\Ev(\vv,t,\kv)\simeq ie \, \kv\T/( \kv\T^2+\gamma^{-2}\kv^2)
\, \exp(-{i{\rm v}k\L t})  \,,
\ee
where ``T'' refers to the transverse components relative to $\vv$. 
Equation (\ref{WW})  describes a cloud of virtual, but ``almost real''  (close to the mass shell) photons. 
This is the basis of the Weizs\"acker-Williams method  \cite{JACKSON}.
If the electron is suddenly deflected or stopped, these photons are ``dropped", forming the {\it initial state bremsstrahlung} (ISB). 
They are lost for the scattered electron, which is temporarily ``half-bare''.  A picture in real space is given in Ref. \cite{SHULGA}. The electron recovers its photon cloud gradually;  
for the component of momentum $\kv$, it needs a distance $\sim \lc(\omega,\gamma,v_\perp)$ 
\cite{FEINBERG,FOMIN-SHULGA,SHULGA-naked,BOLOTOVSKY}. 
Meanwhile, real photons are emitted, forming the {\it final state bremsstrahlung} (FSB). The latter is the same as the the initial state bremsstrahlung of a suddenly stopped {\it positron}, according to the superposition principle (see Fig. 1). 
%
\begin{figure}  [b!]  
\centering
\includegraphics*[angle=90, width=0.70\textwidth, bb= 280 80 420 740]{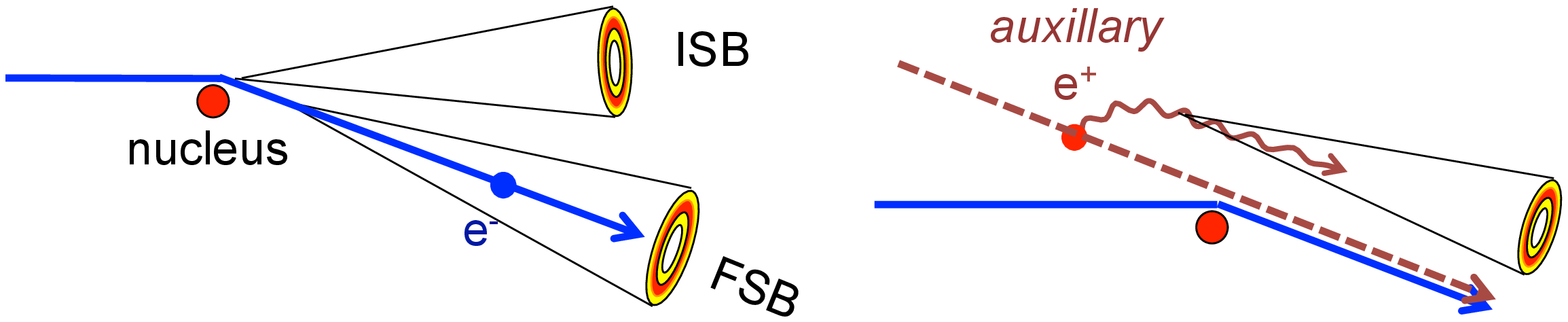}
\caption{Left: initial- and final state bremsstrahlung. Right: equivalence between FSB from an electron and ISB from a positron, according to the superposition principle (the compound $e^+e^-$ system on the right of the nucleus is neutral, therefore does not radiate).}
\end{figure}
%
From  Eq. (\ref{noncov'}), the full bremsstrahlung amplitude at one scattering is 
\be\label{sudden} 
 \arad(\kv)  
=   (ie/\omega) \, e^{i\phi_1} (\appvt_1 -  \appvt_0 ) \,, 
\ee
%
$\phi_1=\omega\td_1$ being the emission phase at the kink. $\appvt_0$ and $\appvt_1$ are the initial and final apparent transverse velocities; they yield the ISB and FSB contributions respectively in Eq. (\ref{sudden}). For a \textit{multiple scattering} event, represented by $N$ kinks, the bremsstrahlung amplitude is 
\be\label{bremsst} 
 \arad(\kv)  
=   \frac{ie}{\omega} \sum_i  e^{i\phi_i} \Delta_i\appvt 
= \frac{ie}{\omega} \left[
e^{i\phi_N} \appvt_N - e^{i\phi_1} \appvt_0 +  
\sum_{j=1}^{N-1} (e^{i\phi_{j+1}} - e^{i\phi_{j}}) \appvt_j 
\right],
\ee
with $\phi_i = \omega\td_i$ and  $\Delta_i\appvt = \appvt_i -  \appvt_{i-1}$. 
For a given $\omega$ but arbitrary $\nv$, if all segments are larger than $\lc(\omega,\gamma,0)=\lambda\gamma^{-2}/\pi$, the phase factors $e^{i\phi_i}$ can be considered as random, therefore one can neglect the interferences between the waves emitted at different kinks. It gives 
\be \label{incoBR} 
\frac{dW}{d\omega d\Omega} = \frac{\alpha}{\pi^2}   \left\langle \sum_i |\Delta_i\appvt|^2 \right\rangle\,,
\ee
which is the incoherent sum of single-scattering bremsstrahlungs. 
This is the case at large enough $\omega$ or in a dilute enough medium. 
Conversely, for a given $\kv$, if a trajectory segment ${\bf L}_j$ is shorter than the coherence length, the ISB at point $j+1$ is suppressed due to the electron undressing. More precisely, it interferes destructively with the FSB at point $j$, since $\phi_{j+1} - \phi_j \equiv L_i/\lc(\omega,\gamma,v_\perp) \ll1$ and the factor $(e^{i\phi_{j+1}} - e^{i\phi_j})$ of Eq. (\ref{bremsst}) is small. This is the root of 
the Landau-Pomeranchuk-Migdal (LPM) and Ternovskij-Shul'ga-Fomin (TSF) effects
\cite{FOMIN-SH-SH}.

The TSF effect is simpler to explain. It occurs in a medium of thickness $L$ large enough to have $\bar\theta_{\rm ms}(L) \gtrsim 1/\gamma$, but less than the typical coherence length $\lc\left(\omega,\gamma,\bar\theta_{\rm ms}(L)\right)$, obtained from Eq. (\ref{Lcoh}) by replacing $\vv\t$ by $\bar\theta_{\rm ms}(L)$, the mean multiple scattering angle. For those $\kv$ such that
\be
\phi_{\rm out} - \phi_{\rm in} \equiv (\omega /2) \int_0^L dt \, [\gamma^{-2} + \vv_\perp^2(t)] \ll 1\,, 
\ee
%
we have $ \arad(\kv) \simeq (ie/\omega) \, e^{i\phi_1} \, (\appvt_{\rm out} - \appvt_{\rm in})$, with $\appvt_{\rm in}\equiv \appvt_0$ and $\appvt_{\rm out}\equiv \appvt_N$, and 
\be \label{mince} 
\frac{dW}{d\omega d\Omega} = \frac{\alpha}{\pi^2} \left\langle  |\appvt_{\rm out} - \appvt_{\rm in}|^2 \right\rangle .
\ee
Equation (\ref{mince}) is to be compared with the incoherent case, Eq. (\ref{incoBR}). Concerning the {\it real} transverse velocities, we have 
\be \label{ivrogne} 
\langle (\vv_{\perp\, \rm out} - \vv_{\perp\,\rm in})^2 \rangle = \left\langle \sum_j |\Delta_j\vv_\perp|^2 \right\rangle ,
\ee
since multiple scattering can be considered as a random walk in transverse velocity space. The same is not granted for the {\it apparent} transverse velocities, since $\appvt$ depends nonlinearly on $\vv_\perp$.
%
Let us consider two cases: 
\begin{description}
\item[a)]  $|\vv_j-\vv_{\rm in}|^2 \ll \gamma^{-2}+ \vv_{\perp\,\rm in}^2$ 
in all segments (dipole regime). In this case the domain of variation of $\vv_\perp$ is small enough to consider $\appvt$ as a linear function of $\vv_\perp$. The analogue of Eq. (\ref{ivrogne}) holds for $\appvt$, then Eq. (\ref{mince}) is equivalent to  Eq. (\ref{incoBR}) {\it as if} there was no interference between successive single-scattering bremsstrahlungs. In fact {\it there is} a destructive interference between FSB at point $j$ and ISB at point $j+1$, but compensated by an average constructive interference between two consecutive FSBs (or two consecutive ISBs). 

\item[b)]  $|\vv_j-\vv_{\rm in}|^2 \gg \gamma^{-2}+ \vv_{\perp\,\rm in}^2$ 
in at least one segment (non-dipole regime). Here we have
%
\be \label{ours} 
\langle (\appvt_{\rm out} - \appvt_{\rm in})^2 \rangle 
< \left\langle \sum_j |\Delta_j\appvt|^2 \right\rangle .
\ee
Indeed, each $\appvt$ is confined in the disk $|\appvt|\le\gamma$, therefore left-hand side of Eq. (\ref{ours}) is bounded by $4\gamma^2$ (see Fig. 2 right), no matter how large is the right-hand side. Then the bremsstrahlung intensity,  Eq. (\ref{mince}) is less than the incoherent case,  Eq. (\ref{incoBR}).%
\footnote{An exception is when $|\vv_{\perp\, \rm out} - \vv_{\perp\,\rm in}|$ comes essentially from {\it one} large-angle single scattering. } 
This is the TSF effect {\it for fixed photon direction}.
If one integrates over $\nv$, the TSF effect appears as a logarithmic, instead of linear, growth of $dW/d\Omega$ with $L$. Indeed, the integral of Eq. (\ref{mince}) over $\nv$, 
\be \label{rapidity} 
{dW}/{d\omega} = 8\alpha \, [\eta/\tanh(\eta)-1] \,,
\ee
where $\gamma( |\vv_{\rm out} - \vv_{\rm in}| = \gamma\theta_{\rm ms} \equiv 2\sinh(\eta/2)$,%
\footnote{$\eta$ is the {\it rapidity} of the outgoing electron in the rest frame of the ingoing one.
}
 increases logarithmically with $\gamma\theta_{\rm ms}$ for $\gamma\theta_{\rm ms}\gg1$, 
while $\theta_{\rm ms} \sim \bar\theta_{\rm ms}(L)$ grows roughly like $\sqrt L$. 

%
\begin{figure} 
\centering
\includegraphics*[angle=90, width=0.70\textwidth, bb= 125 70 470 730]{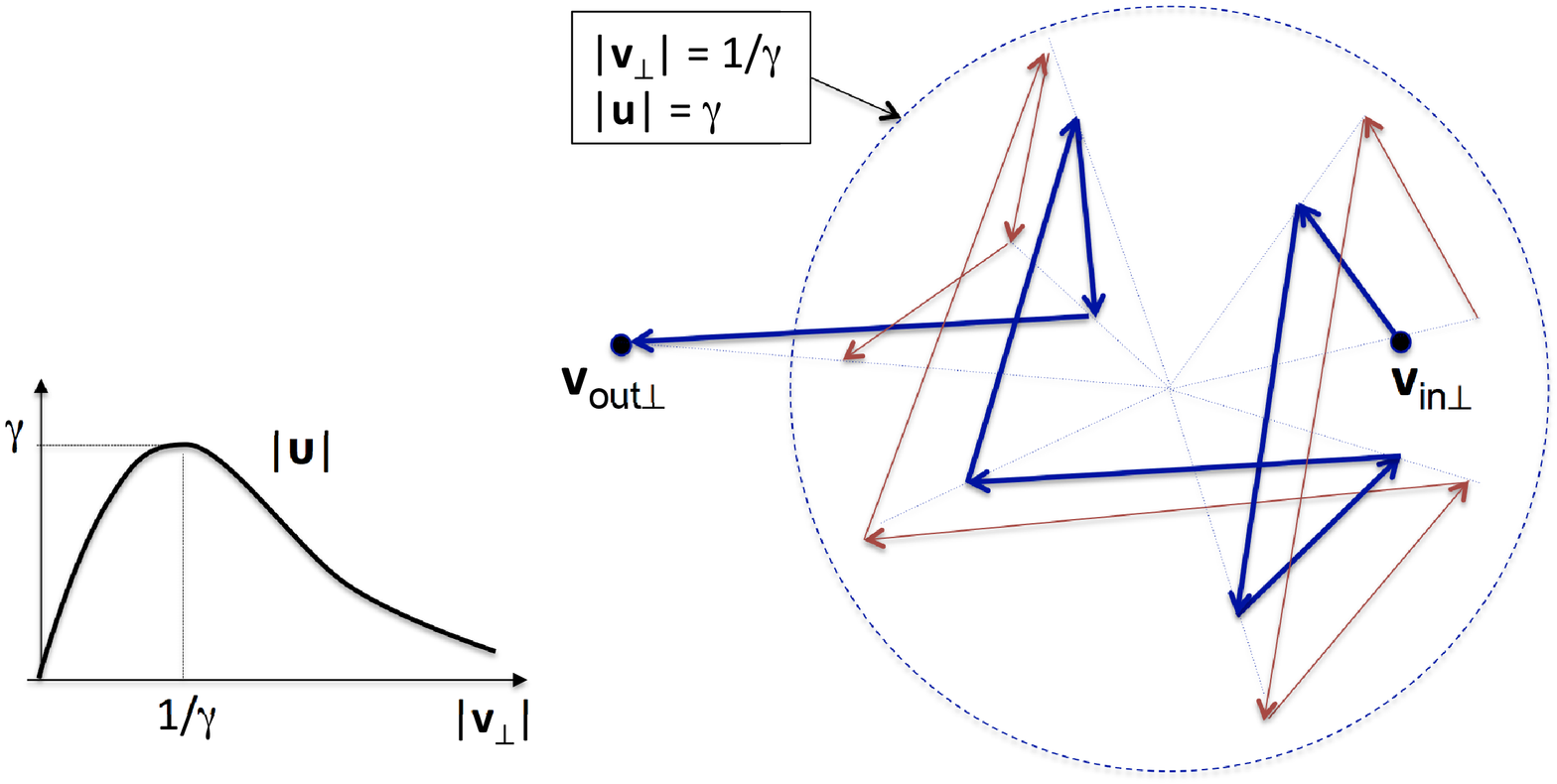}
\caption{Left: apparent versus real transverse velocities. Right:  random walk of $\vv_\perp$ in multiple scattering (thick line) and corresponding walk of $\appvt/\gamma^2$ (thin line). The dotted circle shows the bound on $|\appvt|$.}
\end{figure}
\end{description}

\ni The $LPM$ effect \cite{SHULGA'} is a reduction of bremsstrahlung in a slab of thickness $L>\lc(\omega,\gamma,0)$ 
but in which the multiple scattering angle within a path $\Delta L\sim \lc$ exceeds $\gamma^{-1}$. A phenomenological explanation is the following: Let us divide the slab in sub-slabs of thicknesses $\Delta L$ given by the implicit formula $\Delta L = (\pi/2) \lc\left(\omega,\gamma,\bar\theta_{\rm ms}(\Delta L) \right)$.
In one sub-slab, taken alone, we have $\bar\theta_{\rm ms}(\Delta L) > \gamma^{-1}$ and  
$\Delta\phi \equiv \phi_{\rm out} - \phi_{\rm in} \simeq \Delta L/ \lc \simeq \pi/2$. 
Due to the latter equation, the interferences between successive sub-slabs are, on the average, neither constructive nor destructive. On the other hand $\Delta\phi$ is small enough to allow a TSF effect in each slab. We have therefore a total bremsstrahlung intensity which grows linearly with the number of slabs, \ie, with L, but is suppressed by the TSF effect of the individual slabs. This is the  LPM effect, which is considered here as a {\it multiple TSF effect}. 

\paragraph{Connection with the sum rules.} 
Let us compare two slabs of the same mater and of equal weight per unit area, but one dense, one dilute. They give the same $\bar\theta_{\rm ms}$. Considering electron trajectories with the same set of $\vv_{i\perp}$ in both slabs, Eq. (\ref{sumrule}c) yields formally the same classical moment $M^{(1)}(\epol,\nv)$, ignoring the ultraviolet divergence. On the other hand, Eqs. (\ref{sumrule}a,b) yield formally (ignoring infrared divergences) smaller $M^{(-3)}(\epol,\nv)$ and $M^{(-1)}(\epol,\nv)$ in the dense slab than in the dilute one, due to the shorter paths between two scatterings. Thus, there is a priori less soft photons (and more hard photons) emitted in the dense slab, in accordance with the LPM and TSF effects. However this reasoning erroneously predicts that these effects also occur at $\gamma\theta_{\rm ms}\ll1$. 
A proper reasoning should use filtered sum rules eliminating the infrared and ultraviolet divergences.

\section{Summary}

Expressing the classical radiation amplitude in terms of the apparent transverse velocity, we have formulated spectral sum rules, linearly of circularly polarized, and indicated the way to make them convergent.  These sum rules can be used to test computing codes of radiation sources like channeling radiation and coherent bremsstrahlung.  
We have also made the connection between these sum rules and atomic sum rules for the oscillator strengths. Then we have revisited the TSF and LPM effects. Their fundamental origin is the non-linear dependence of the apparent transverse velocity versus the real transverse velocity. 


\begin{thebibliography}{99}



\bibitem{B-K} V. N. Baier and V. M. Katkov, Zh. Eksp. Teor. Fiz. 
\textbf{53} (1967), 1478 and
\textbf{55} (1968), 1542 [Sov. Phys. JETP \textbf{26} (1968), 854 and \textbf{28} (1969), 807].

 \bibitem{XA-Ischia} X. Artru, {\it Radiation by an electron in an electromagnetic plane wave: The quasiclassical formula and other domains of applications}, Phys. Rev. Accel. and Beams {\bf 22} (2019) 050705.

\bibitem{JACKSON} J. D. Jackson, {\it Classical Electrodynamics} (Wiley, N.Y.,
1999).

\bibitem{RAD} 
P.  Rullhusen, X. Artru and P. Dhez, {\it Novel Radiation Sources Using Relativistic Electrons} (World Scientific, 1998).

\bibitem{XA-spont-stim} X. Artru, {\it Spontaneous and stimulated radiations of a classical moving charge from energy conservation}, Nucl. Instr. and Meth. B {\bf 402} (2017) 112.

\bibitem{GOUANERE} M. Gouan\`ere, D. Sillou, M. Spighel, N. Cue, M. J. Gaillard, R. G. Kirsch, J.-C. Poizat, J. Remillieux, B. L. Berman, P. Catillon, L. Roussel and G. M. Temmer, Phys. Rev. B {\bf 38} (1988), 4352.

\bibitem{XA-Elbruz} X. Artru, {\it Sum rules for synchrotron radiation in nonuniform field. Application to channeling radiation}, Radiation Effects and Defects in Solids, {\bf116} (1991), 293.

\bibitem{XA-code} X. Artru, {\it A simulation code for channeling radiation by ultrarelativistic electrons or positrons,} Nucl. Instr. and Meth. B {\bf 48} (1990) 278.

\bibitem{Garnier} P. Garnier, Academic training, 1998 (unpublished).

\bibitem{Bethe} H.A. Bethe and E.E. Salpeter, {\it Quantum mechanics of one- and two-electron systems}, Handbuch der Physik {\bf35} (Springer 1957)

\bibitem{Klar} H. Klar, Phys. Rev. Lett. {\bf59} (1987) 1656; Z. Phys. {\bf D8} (1988) 261.

\bibitem{SHULGA} A. I. Akhiezer and N. F. Shul'ga, 
{\it High-Energy Electrodynamics in Matter}, Gordon and Breach, 1996.

\bibitem{FEINBERG} 
E.L. Feinberg, {\it Hadron clusters and half-dressed particles in quantum field theory}, Sov. Phys. Uspekhi {\bf 23} (1980) 629. 

\bibitem{FOMIN-SHULGA} S. P. Fomin and N.F. Shul'ga, 
{\it On the space-time evolution of the process of ultrarelativistic electron radiation in a thin layer of substance}, 
Phys. Lett. A {\bf 114} (1986), 148. 

\bibitem{SHULGA-naked} N.F. Shul'ga and V.V. Syshchenko, J. Phys. Atom. Nucl. {\bf 63} (2000), 11. 

\bibitem{BOLOTOVSKY} B. M. Bolotovskii and A.V. Serov, Pis'ma Zh. Eksp. Teor. Fiz. {\bf90} (2006), 468. 

\bibitem{FOMIN-SH-SH} 
N.F. Shul'ga and S.P. Fomin, JETP Lett. {\bf 27} (1978) 117;  JETP {\bf 86} (1998) 32;  {\it Suppression of radiation in an amorphous medium and in a crystal}, Nucl. Instr. and Meth. B {\bf 145} (1998) 73.

\bibitem{SHULGA'}
A.I. Akhiezer, N.F. Shul'ga and S.P. Fomin, {\it Landau-Pomeranchuk-Migdal Effect}, Physics Reviews {\bf 22}, Part 1, ed. by I. M. Khalatnikov, Cambridge Scientific Publishers, 2005.


 
 


\end{thebibliography}
\end{document}